\documentclass[aps,twocolumn,prd,showpacs,nofootinbib,showkeys,superscriptaddress,preprintnumbers]{revtex4-1}
\usepackage{hyperref}
\usepackage{graphicx,floatflt}
\usepackage{amsmath}
\usepackage{amstext}
\usepackage{epsfig}
\usepackage[usenames]{color}
\usepackage[dvipsnames]{xcolor}

\newcommand{\dd}{\mathrm{d}} % For the derivatives
\newcommand{\rr}{\mathrm}

\newcommand{\Msun}{M_\odot}

\newcommand{\del}{\delta_{\rm PBH}}%^{\rm local}}

\newcommand{\mPBH}{m_{\rr{\rm PBH}}}

\newcommand{\be}{\begin{equation}}
\newcommand{\ee}{\end{equation}}
\newcommand{\ba}{\begin{eqnarray}}
\newcommand{\ea}{\end{eqnarray}}

\begin{document}

%\preprint{IFT-UAM/CSIC-17-108, CERN-TH-2017-239}

\title{Seven Hints for Primordial Black Hole Dark Matter}

\author{S\'ebastien Clesse}
\email{sclesse@uclouvain.be, sebastien.clesse@unamur.be}
\affiliation{Centre for Cosmology, Particle Physics and Phenomenology (CP3), Institut de Recherche en Mathematique et Physique (IRMP), Louvain University, 2 Chemin du Cyclotron, 1348 Louvain-la-Neuve, Belgium}
\affiliation{Namur Center of Complex Systems (naXys), Department of Mathematics, University of Namur, Rempart de la Vierge 8, 5000 Namur, Belgium}

\author{Juan Garc\'ia-Bellido} \email{juan.garciabellido@cern.ch} 
\affiliation{Instituto de F\'isica Te\'orica UAM-CSIC, Universidad
Auton\'oma de Madrid, Cantoblanco, 28049 Madrid, Spain}
\affiliation{CERN, Theoretical Physics Department, 1211 Geneva, Switzerland}

\date{\today}

\begin{abstract}
Seven observations point towards the existence of primordial black holes (PBH), constituting the whole or an important fraction of the dark matter in the Universe:  the mass and spin of black holes detected by Advanced LIGO/VIRGO, the detection of micro-lensing events of distant quasars and stars in M31, the non-detection of ultra-faint dwarf satellite galaxies with radius below 15 parsecs, evidences for core galactic dark matter profiles, the correlation between X-ray and infrared cosmic backgrounds, and the existence of super-massive black holes very early in the Universe's history.    Some of these hints are newly identified and they are all intriguingly compatible with the re-constructed broad PBH mass distribution from LIGO events, peaking on PBH mass $m_{\rm PBH} \approx 3 M_\odot$ and passing all other constraints on PBH abundances.   PBH dark matter also provides a new mechanism to explain the mass-to-light ratios of dwarf galaxies, including the recent detection of a diffuse galaxy not dominated by dark matter.   Finally we conjecture that between 0.1\% and 1\% of the events detected by LIGO will involve a PBH with a mass below the Chandrasekhar mass, which would unambiguously prove the existence of PBH.

\end{abstract}
%\pacs{98.80.Cq}fC

\maketitle

Despite a series of indirect evidences the nature of Dark Matter (DM) accounting for about 85\% of today's matter density of the Universe  remains a mystery.  The failure of a long series of direct detection experiments~\cite{Cui:2017nnn,Aprile:2017iyp,Akerib:2016vxi} to detect a Weakly Interacting Massive Particle (WIMP) or any other particle candidate now leads some cosmologists to reconsider alternatives to particle Dark Matter models.   One of them has received recently a lot of attention, after the detection of several black hole mergers by Advanced LIGO/VIRGO since September 2015~\cite{Abbott:2016blz,Abbott:2016nmj,Abbott:2017vtc,Abbott:2017oio}:  the possibility that Dark Matter could be constituted in part or entirely by Primordial Black Holes (PBH).   

PBH are naturally an ideal Dark Matter candidate, being non-luminous, non-relativistic and nearly collisionless.    They could have formed in the early Universe due to the gravitational collapse of pre-existing order one density fluctuations, eventually generated during inflation.
%~\cite{GarciaBellido:1996qt,2011arXiv1107.1681L,Bugaev:2011wy,Clesse:2015wea,Garcia-Bellido:2016dkw,Domcke:2017fix,Garcia-Bellido:2017aan,Garcia-Bellido:2017mdw,Ezquiaga:2017fvi,Jedamzik:1999am,Rubin:2001yw,Kohri:2012yw,Kawasaki:2012wr,Cotner:2016cvr,Deng:2016vzb}.  
 The unexpected large mass of the GW150914 and GW170814 black holes~\cite{Abbott:2016blz,Abbott:2017oio}, combined with the inferred merger rate coinciding with PBH abundances comparable to the one of Dark Matter~\cite{Bird:2016dcv,Clesse:2016vqa}
has revived the interest for PBH as a plausible Dark Matter candidate.   Since then, a series of new studies have established improved constraints in the mass range [10-100] $\Msun$, e.g. from the dynamics and the existence of star clusters in faint dwarf galaxies~\cite{Brandt:2016aco,Green:2016xgy,Li:2016utv,Koushiappas:2017chw} and from the non-observation of X-ray and radio sources towards the galactic center~\cite{Gaggero:2016dpq}.  Others have re-evaluated the constraints from the CMB temperature anisotropies~\cite{Ali-Haimoud:2016mbv,Blum:2016cjs,Poulin:2017bwe}, which are indirectly affected by matter accretion onto PBH that impacts on the thermal history of the Universe.  When these constraints are combined with the ones coming from the non-observation of microlensing of stars in the Magellanic clouds~\cite{Tisserand:2006zx}, it seems that the whole range [1-100] $\Msun$ is now excluded.  

In this \textit{letter}, we argue that a careful investigation of these constraints reveals that the case of a broad mass distribution in the range  $[1-10] \Msun$ is still allowed.   By using the latest LIGO merging events and rates, we reconstruct the plausible PBH mass spectrum, in case they constitute the dominant component of Dark Matter, and find that it falls exactly within the allowed range.   Even more, by reanalyzing previous constraints and by using physical arguments based on first principles, we show that there exists seven types of observations pointing towards this PBH-DM scenario~\cite{Garcia-Bellido:2017fdg}.  Finally, we claim that PBH as cold dark matter would provide a natural explanation to the longstanding problems of cosmology and astrophysics, namely the existence of supermassive black holes at high redshifts, the missing galactic satellites, the too-big-to-fail, the core-cusp problems, and the missing baryons in the Universe. 

{\bf \textit{Hint 1}:  Black Hole merger rates and reconstruction of the mass spectrum.}   One can calculate the merger rates expected if PBH constitute all the dark matter, following and extending~\cite{Bird:2016dcv,Clesse:2016vqa}.   These rates depend on the shape of the PBH mass spectrum.  We consider a PBH-DM model with a log-normal density distribution~\cite{Clesse:2015wea} of width $\sigma$, centered on the mass $\mu$, but our results can easily be extended to any mass function and any PBH abundance.  Such a lognormal spectrum is motivated by PBH formation models in the context of inflation, e.g. in mild-waterfall hybrid inflation~\cite{Clesse:2015wea}.  The merger rate of PBH binaries strongly depends on how PBH cluster.  Two models have been considered:  

\textit{1. Dominant clustering scale (DCS):}   The majority of BH merging events occurred in PBH clusters of similar sizes and concentration.   One can encode together the typical clustering scale and the halo profile uncertainties in a single parameter $\langle \del^2 \rangle^{1/2}$~\cite{Clesse:2016vqa,Clesse:2015wea} representing the r.m.s of the PBH density contrast (with respect to the DM cosmological density) within these halos.   Typically $\langle \del^2 \rangle^{1/2} $ could vary from $\mathcal O(10^7) $ for the largest dwarf spheroidals, up to $\mathcal O(10^{11})$ for dense globular clusters.    %Larger values could be obtained for extremely dense micro-clusters (e.g. of milli-parsec size) in which two-body interactions are frequent and able to disrupt early-formed binaries.  

\textit{2. Extended halo mass function (Ext-HMF):}   Instead of assuming a natural clustering scale, one can consider an extended halo mass function such as the one expected from the standard merger-tree paradigm of structure formation.   Analytical estimates from the Press-Schechter formalism or low-mass extrapolations of N-body simulations have been used in~\cite{Bird:2016dcv} to compute the PBH merger rate for a monochromatic mass distribution.   We have extended their calculation to the case of a broad spectrum. 

The two models naturally induce PBH merger rates consistent with the ones inferred by LIGO, from a few to a few hundreds events/year/Gpc$^3$ (details on the rates are provided in Appendix~\ref{sec:rates}).   This coincidence, combined with the mostly unexpected large mass of the BH progenitors, has been the first hint pointing to the existence of a PBH-DM population.  

\begin{figure*}
\begin{center}
\includegraphics[scale=0.62]{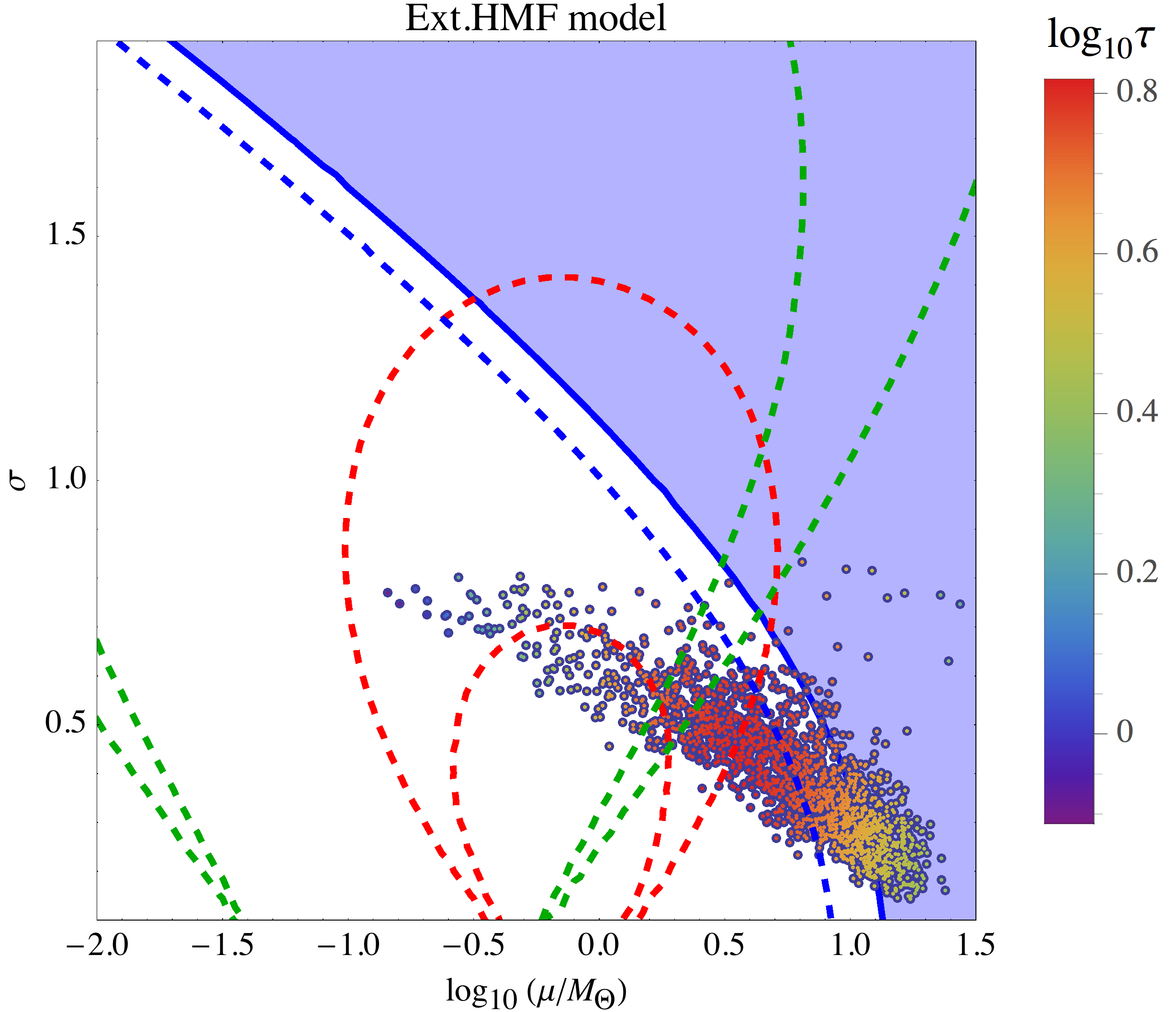} ~
\includegraphics[scale=0.62]{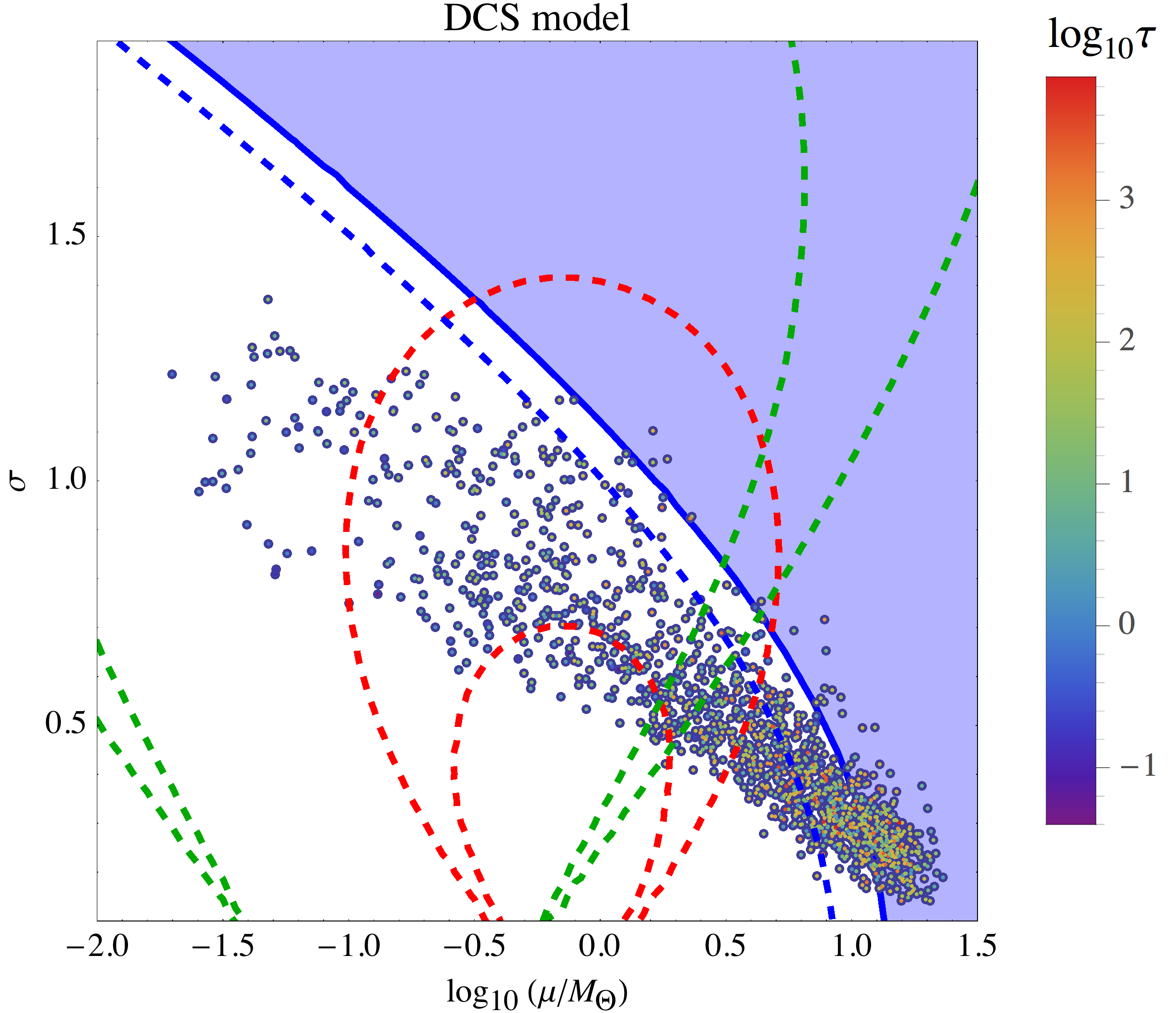}
\caption{Reconstruction of the PBH mass spectrum assuming a log-normal density distribution of mean $\mu$ and width $\sigma$, for the extended halo mass function (Ext.HMF, left panel) and the dominant clustering scale (DCS, right panel) models.  The color scale of the MCMC points represent the model merger rates in units of $\rm{Gpc^{-3} yr^{-1}}$.    The shaded blue region is the parameter space excluded by the dynamical heating of star clusters in the faint dwarf galaxy  Eridanus II, assuming it has a central $6500 M_\odot$ IMBH (solid blue), or not (dashed blue).   M31 and quasar microlensing events favor the regions between the red and green dashed lines respectively.}
\label{fig:spectrumreconstruction}
\end{center}
\end{figure*}

\begin{figure}
\begin{center}
\includegraphics[scale=0.45]{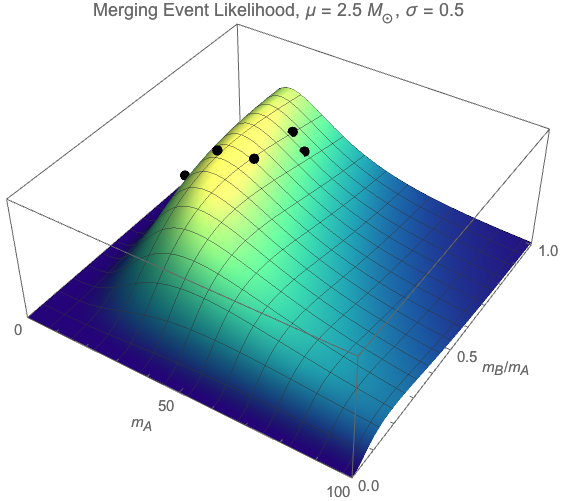}
\caption{Single merging event likelihood for a PBH model with a braod log-normal density spectrum with central value $\mu = 2.5 \Msun$ and width $\sigma = 0.5$.  The best-fit BH masses of LIGO merging events are represented by the black points. }
\label{fig:merglik}
\end{center}
\end{figure}

What are the preferred values of the mass distribution parameters $\mu$ and $\sigma$ given the five GW events detected so far?   Starting from the black hole progenitor masses $m_i$ and the inferred merger rate $\tau$ as well as their uncertainties,  we have reconstructed the Bayesian posterior probability distribution of $\mu$ and $\sigma$, using a Markov-Chain Monte-Carlo (MCMC) method.   The MCMC points are represented in Fig.~\ref{fig:spectrumreconstruction} for the two clustering models.  The color scale indicates the merger rate expected for LIGO.   Because LIGO more likely detects heavier BH in the tail of the mass distribution, the preferred value of $\mu $ ranges from $\sim 0.1 (0.01)$ up to about $\sim 20 \Msun$ and $ 0.1 \lesssim \sigma \lesssim 0.8 (1.0)$ for the Ext-HMF (DCS) model.    The LIGO event likelihood obtained for $\mu = 2.5 \Msun$, $\sigma = 0.5$ is represented in Fig.~\ref{fig:merglik}.  It peaks on masses about four times larger than $\mu$ and on mass ratios down to $m_{\rm B}/m_{\rm A}  \approx 0.2$.   One would therefore expect Adv\-LIGO to be able to detect a subdominant fraction (between 0.1\% and 1\% for the preferred scenarios) of PBH mergers involving a PBH of mass $m_{\rm A} \simeq 5\,\Msun$ and a PBH with a mass smaller than the Chandrasekhar mass, $m_{\rm B} \lesssim M_{\rm Ch} \simeq 1.4\,\Msun$, which would be a clear indication of primordial origin. These events have chirp masses $M_c = (m_{\rm A}\,m_{\rm B})^{3/5}/(m_{\rm A}+m_{\rm B})^{1/5} \simeq 1.8\,\Msun$, which are too faint to be detected by the next run O3 of AdvLIGO-Virgo, unless they are within 200 Mpc of Earth.

{\bf \textit{Hint 2}: Black hole spins.}  Because PBH formed due to the collapse of (nearly) Gaussian overdensities in the early Universe, they have initially low spins~\cite{Chiba:2017rvs} that can only be enhanced by very rare mergers or by matter accretion.  In the case of PBH-DM, it is expected that a limited fraction of PBH initially formed short-lived binaries and the remaining black hole after their merging would have an enhanced spin $ \chi_{\rm eff} = (m_{\rm A}\chi_{\rm A} + m_{\rm B}\chi_{\rm B})/M = (m_{\rm A}\vec a_{\rm A}\cdot\vec L + m_{\rm B}\vec{a}_{\rm B}\cdot\vec L)/(M|\vec L|) \simeq 0.7$~\cite{Fishbach:2017dwv}, with $\vec{a}_{\rm A,B} \equiv c \vec S_{\rm A,B}/ G m_{\rm A,B}^2$, the black hole spins $ \vec S_{\rm A,B} $ and the orbital momentum $\vec L$.  But one also expects PBH in dense DM-dominated halos (with typical halo mass below $\sim 10^9 \Msun$ as argued thereafter) to have accreted most of the baryonic matter.    The accretion should be isotropic due to the random BH motion with respect to the baryons, which should reduce the BH spin down to $\chi_{\rm eff} \lesssim 0.1$~\cite{Berti:2008af}.   If instead BH have a stellar origin, conservation of angular momentum typically implies large spin values after contraction, eventually sped up by a subsequent accretion phase of a companion star before it explodes and becomes a BH~\cite{Farr:2017uvj}.   Instead of taking a random direction, the BH from a pre-existing star binary would tend to align their spin with the orbital angular momentum~\cite{Belczynski:2017gds}. In the simplest scenarios, one would therefore expect $\chi_{\rm eff} \gtrsim 0.8$, whereas LIGO mergers have effective spins centered on $\chi_{\rm eff} \simeq 0$ (specially for GW170814~\cite{Abbott:2017oio}).

In June 2017, LIGO announced the detection of GW170104~\cite{Abbott:2017vtc}.  It has been possible to reconstruct with good confidence the spin orientation of the heaviest black hole.  It has been found to be anti-aligned with the orbital angular momentum and this led LIGO to claim for the discovery of a new population of black holes forming binaries through a capture process, just as expected for PBH.   A viable competing scenario is the one of stellar BH populations in very dense environments such as the core of globular clusters or nearby supermassive black holes.  However it is still unclear whether this scenario can combine both an initially low-metallicity environment for massive stellar BH to form and the existence of abundant very dense environments with  black holes close enough to explain the merger rates inferred by LIGO.   
%Regarding the spins, if those BH were formed isolated from a stellar explosion, their spin should be quite low.   This is therefore in contradiction with the most likely total spin of GW150914 and the heaviest BH spin of GW170104, the two other merger events being nevertheless consistent with $\chi_{\rm eff} = 0$.   
On the other hand, if an exotic scenario could explain BH spin misalignment for a stellar binary origin, it would be difficult to explain why all the LIGO mergers detected so far have total spins  $\chi_{\rm eff} \lesssim 0.5$, whereas the spins in high-mass X-ray binaries go up to $\chi \simeq 0.95$, in opposition with  the limits imposed by LIGO on the spin of the most massive progenitors $a_1 < 0.7$ for GW150914 and LVT151012 and $a_1 < 0.8$ for GW151226 at 90\% C.L., large parallel spins aligned with the orbital momentum being disfavored~\cite{TheLIGOScientific:2016pea}. 

Putting all together, the PBH DM scenario seems to be in agreement with the spins and rates observed in the LIGO BHB mergers, which most probably have a capture origin.  Observations are thus challenging the previously widely accepted scenario of a stellar binary origin~\cite{Farr:2017uvj}.  

% Future observations will help to distinguish PBH from the competing scenario of stellar BH binary formations in very high density environments.   Ultimately, the detection of the stochastic gravitational wave background by LISA will be able to distinguish between the two models, because different density environments impact the shape of the GW spectrum on frequencies $f \lesssim 10^{-4}$ Hz~\cite{Clesse:2016ajp}.
 
{\bf \textit{Hint 3:} Microlensing of stars in M31 and distant quasars. }   Microlensing of stars is the best possible way to detect the presence of PBH in galactic halos.  Contrary to surveys of the LMC and SMC that probe only a tiny fraction of the Milky Way halo, surveys of M31 have monitored stars spread over the whole galactic disk and bulge, and are thus a probe of an important fraction of the Andromeda dark matter halo.  The expected number of events is thus not suppressed by a low-probability to find PBH halos along the lines of sight.  Up to date a total of 56 microlensing events have been detected in M31, several of them by different instruments~\cite{2015ApJ...806..161L}.  The timescale goes from short (few days) to very long (50-100 days) events.  Some of them might be attributed to intrinsic varying disk stars or long period variables.  But numerous events are very unlikely to be due to self-lensing of stars.  This therefore points towards the presence of massive compact objects in the M31 halo.  Their abundance is still rather poorly constrained, but studies~\cite{CalchiNovati:2005cd}  have claimed a non-zero fraction with a lower limit of about 20\% if the object mass is within [0.5-1] solar mass at 95\% C.L, and upper bound of about 100 solar masses.  

Microlensing of distant quasars is another way to probe the MACHO content of galactic halos.  Very recently, microlensing from 24 gravitational lensed quasars in optical and X-ay have been analyzed~\cite{Mediavilla:2017bok}.  It results that about $20 \pm 5 \%$ of the total matter in the galaxy lens is made of any type of compact objects with masses in the range $0.05 \Msun \lesssim M \lesssim 0.45$.   Such abundance is in strong tension with the expected stellar component within the lens galaxies, and so these observations point towards an important MACHO component. 

M31 and quasar bounds on the PBH content are in apparent contradiction with the constraints from the EROS survey towards the LMC and SMC.   But EROS bounds have been recently reanalyzed and are now considered as much less stringent than initially claimed~\cite{Hawkins:2015uja,Green:2017qoa,Garcia-Bellido:2017xvr}.  The consistency  between different microlensing surveys is further discussed in Appendix~\ref{sec:microlensing}.   The window for PBH dark matter in the solar mass range has been re-opened.

We have represented in Fig.~\ref{fig:spectrumreconstruction} the regions in the $(\mu, \sigma)$ plane leading to $20-35$ \% of PBH in the range $[0.5-1] M_\odot $ and to $15-25$ \% in the range $[0.05-0.45] M_\odot$, as suggested by M31 and quasar microlensing.   These two regions overlap for $\mu \sim 3 \Msun  $ and $\sigma \sim 0.6$, which intriguingly exactly falls within the ellipse obtained by the mass spectrum reconstruction with LIGO events.   

{\bf \textit{Hint 4}:  Dynamics and star clusters of ultra-faint-dwarf-galaxies.}    The recent detection of numerous satellite ultra-faint dwarf galaxies (UFDG) of our galaxy, in M31 or in the Local Group, actually provide not only one but several clues of the PBH-DM existence.   Until now they have been used, though, only to narrow the viable PBH masses and abundances. The most stringent constraint come from the existence of a central cluster in Eridanus II~\cite{Brandt:2016aco,Green:2016xgy,Li:2016utv} that should have been dynamically heated by the more massive PBH for $\mu \gtrsim 10 \Msun$, unless there is an intermediate mass BH at the centre of these dwarf spheroidals~\cite{Li:2016utv}.  It has been represented in Fig.~\ref{fig:spectrumreconstruction} and is consistent with the mass spectrum reconstruction.  The dynamical heating of stars in UFDG sets less stringent but nevertheless strong constraints on PBH abundances~\cite{Brandt:2016aco,Green:2016xgy}\footnote{More stringent ones were found in~\cite{Koushiappas:2017chw} from the radial profile of the projected surface luminosity of Segues I.  However this cannot be extended to a broad distribution in a simple way, because heavy PBH should not only heat up the stars but also the lighter PBH, an effect that was not considered in~\cite{Carr:2017jsz}}. 

\begin{figure*}
\begin{center}
\includegraphics[scale=0.5]{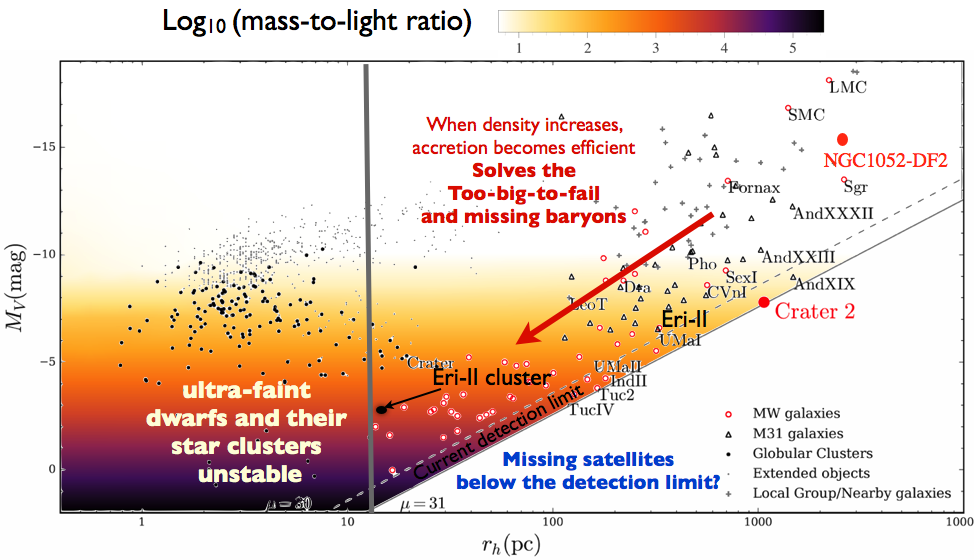}
\caption{Absolute magnitude $M_V$ vs. the half-light radius $r_h$ of dwarf galaxies, globular clusters and extended objects in the Milky-Way (MW), M31 or galaxies of the local group.  The vertical gray line represents our estimation of the limit below which DM-dominated faint dwarf galaxies, or their star cluster, have been dynamically heated by solar-mass PBH, if they constitute all the DM.  The non-detection of faint dwarfs on the left-hand side of this line is a clue in favor of the PBH-DM scenario.   In this scenario, early matter accretion makes dwarf galaxies naturally faint, most of them being potentially located below the detection limit, providing a solution to the missing satellites problem and to the missing baryons.   Accretion onto PBH in large and dense halos would also suppress star formation, which might also explain the too-big-to-fail problem.  The predicted %remaining baryon fraction 
mass-to-light ratio after a PBH accretion episode (for the toy-model discussed in the text) is represented by the colored scale and is roughly consistent with the ones of dwarf galaxies.   The DM dominated Crater 2 and the recently detected diffuse galaxy NGC1052-DF2 lacking of DM, are pointed out and fall respectively below and above the expected transition in mass-to-light ratios.  Figure adapted from~\cite{2016MNRAS.459.2370T}. }
\label{fig:dwarfs}
\end{center}
\end{figure*}

By using the same method of~\cite{Brandt:2016aco,Green:2016xgy} we have computed the critical radius below which the stars in a DM dominated halo are dynamically heated in a timescale of the order of the age of the oldest stars (10 billion years) in UFDG, eventually taking into account the influence of a central IMBH as in~\cite{Li:2016utv}.  Depending on the model parameters, one gets a critical radius between 10 and 20 pc for strongly DM-dominated objects independently of the total halo mass, as represented in Fig.~\ref{fig:dwarfs}.  This intriguingly coincides with the minimal radius below which there has been no observation of UFDG, and of star clusters in UFDG, whereas such objects would still be above the detection limit and should therefore have been detected.    This is a first clue for PBH-DM in UFDG.   On the other hand we have checked that non DM dominated objects, such as globular clusters, are stable.  % for models of Fig.~\ref{fig:spectrumreconstruction}.

On the other hand, because UFDG are strongly DM dominated they are the ideal objects to probe the (particle) DM properties.   In particular, the stability of the central star cluster of Eri II has been investigated.  This cluster is fragile and should not have survived to the strong tidal fields of a central DM density cusp and a scenario where this cluster would have sank to the galactic center is extremely fine-tuned~\cite{Contenta:2017jph,Amorisco:2017hac}.   If the cluster is however long-lived in a cored DM halo, because in UFDG stellar feedback is unlikely to be effective, it therefore points towards another dynamical heating process of the cusp, a second clue of the PBH-DM nature.  

A third clue for PBH-DM comes from the lack of luminous dwarf galaxies, especially the largest ones, compared to expectations from cosmological N-body simulations, two long-standing problems referred as the missing satellite and too-big-to-fail problems.   But it has been recently shown that the inferred number of UFDG, when extrapolating the number of detections to the whole galactic halo, is compatible with the expected number of satellite galaxies~\cite{Kim:2017iwr}.   So there is no missing satellites problem, but rather a faintness problem, i.e. why star formation was inefficient in satellite galaxies?    
PBH could naturally explain why most dwarf galaxies are faint DM dominated objects, due to the early accretion of most of the baryonic matter, thus preventing important star formation.   Using first principles, one can evaluate the baryonic fraction remaining after one billion years, assuming the Hoyle--Bondi mass accretion rate.  We have computed (see Appendix~\ref{sec:accretion} for the detailed calculation) the remaining Baryon fraction after one Gyr for a simple toy model of uniform spherical halos whose mass, radius and Virial velocity relations are calibrated on the half-light radius, Virial velocity and mass-to-light ratio of Eridanus II from \cite{Li:2016utv}, and rescaled using the Virial theorem.  We also assumed that all remaining baryons formed stars.   Our estimation of the resulting mass-to-light ratio is represented in Fig.~\ref{fig:dwarfs}.  It exhibits a transition between luminous and faint dwarfs that roughly coincides with the enhancement of the mass-to-light ratios from observations.   This scenario would also explain why there is no evidence for intermediate-age populations in ultra-faint dwarfs, and why they are composed of old metal poor population II stars that were formed within only one Gyr, which indicates a truncated star formation due to a global event~\cite{2012ApJ...753L..21B}, here a rapid accretion episode.   At same halo mass, a reduction of radius boosts the efficiency of accretion.  This effect should prevent star formation, even in the most massive and dense halos, thus providing a solution to the too-big-to-fail problem.    Notice, however, that the exact position of this transition is model depended and could be shifted easily top or bottom for other choices of parameters and model.  So further investigations are needed to compute more realistically the efficiency of accretion, taking into account more complex effects such as inhomogeneous accretion due to the PBH-DM profile and mass distribution, the formation and evolution of stars, etc.   It is nevertheless encouraging that observations of mass-to-light ratios can be reproduced, at least qualitatively, in the PBH-DM scenario with such a simple toy model.   In particular, PBH dark matter would explain the existence of a recently detected diffuse galaxy lacking of dark matter~\cite{vanDokkum:2018vup}.  With a radius of $~2.2$ kpc and a mass-to-light ratio of at most a few, it is not dense enough for PBH accretion to be efficient and so it differs from other dwarf galaxies like Crater 2 that are strongly DM dominated.  These two galaxies are pointed out on Fig.~\ref{fig:dwarfs}.   Finally, this scenario would also explain the missing baryons problem since at the end of the accretion episode, a potentially important fraction of the baryonic matter of the Universe have fed PBH.

{ \bf \textit{Hint 5}:  Core galaxy profiles from PBH scattering.}   A long-standing problem of the standard cosmological scenario is that N-body simulations of structure formation result in cuspy DM halo profiles, well fitted by a Navarro--Frenck--White (NFW) profile~\cite{Navarro:1996gj}, whereas the halo reconstruction from kinematical properties suggest that dwarf galaxies have a core profile~\cite{deBlok:2009sp}.  Baryonic feedback is a possible explanation but the importance of the effect is still debated~\cite{Weinberg:2013aya}.   In particular, baryonic feedback would not homogenize the DM cusp in the largest halos, neither in low-mass ultra-faint dwarf galaxies, in which star formation in suppressed.  But some recent observations support the existence of a core in halos with those smallest and largest mass scales:  on the one hand, as already discussed, the star cluster at the center of some ultra-faint dwarf galaxies should have been disrupted by the DM cusp and, on the other hand, observations support an offset of 10 to 20 kpc between some brightest cluster galaxies and the center of very massive ($>10^{13} M_\odot$) DM halos~\cite{Harvey:2017afv}.  The galactic wobbling around the DM halo center supports a core profile that would hardly be explained by baryonic feedback.

An alternative scenario is the one of interacting particle dark matter~\cite{Vogelsberger:2015gpr,Cyr-Racine:2015ihg}.   For cross sections between 0.1 and 2 $\rm{cm}^2/\rm{g}$, the kinetic energy exchange between DM particles would induce a core profile compatible with observations~\cite{Brinckmann:2017uve}.  Strangely, gravitational scattering between solar-mass PBH actually leads to a similar cross-section, given by~\cite{Mouri:2002mc}
\be
\frac{\sigma_{\rm PBH} }{m_{\rm PBH}}\approx 0.3 \left[ \frac{m_{\rm PBH}}  { M_\odot} \right] \left[ \frac{v_{\rm rel}}{\rm km/s} \right]^{-4} {\rm cm^2/g}~,
\ee  
if all PBH have the same mass and relative velocities typical to the one expected in the core of some DM halos, $v_{\rm rel} \sim \mathcal O(1) \rm{km/s}$.   In reality, the numerous light PBH from an extended mass distribution could significantly boost the effective cross-section so that larger velocities could be accommodated.  One can determine the typical radius below which gravitational PBH scattering leads to an homogenization of the kinetic energy, and thus of the density, we have computed the relaxation time of the system, assuming a NFW profile initially, $ t_{\rm relax} (r)  \approx  r N /( 8 v_{\rm vir} \ln N) $, where $N$ is the number of PBH within a sphere of radius $r$, and we have determined the typical radius for which homogenization takes place in a time of the order of 10 Gyr.   We assumed for simplicity that all the PBH have the same mass $M_{\rm c}$\footnote{For a broad mass distribution as the one represented in Fig.~\ref{fig:spectrumreconstruction}, with $\sigma \simeq 0.5$ the lightest PBH should be rapidly slingshot away from the halo so that the number of PBH per halo, and thus the core radius, should not vary by more than a factor unity compared to the monochromatic case}.  One obtains a core radius of the order of the kpc for $10^9 M_\odot$ dwarf galaxies, compatible with observations, going down to a few tens of parsecs for ultra-faint dwarf galaxies, and eventually up to $\sim10$ kpc for the most massive DM halos ($10^{14} M_\odot$), which would therefore also explain the wobbling of the brightest cluster galaxies.  
 
 A refined analysis will have to compute the scattering between PBH of different masses, including the effect of a central SMBH or IMBH, and the slingshot of light PBH, which would require the use of N-body simulations.   It is nevertheless remarkable that based on first principles, one gets a possible simple and natural explanation to the core-cusp problem with solar-mass PBH-DM.

{ \bf \textit{Hint 6}: Correlations between X-ray and infrared cosmic backgrounds. }  The massive PBH DM scenario~\cite{Garcia-Bellido:2017fdg} predicts that the tails of the PBH mass distribution will correspond to very massive BH that will act as seeds for gas to fall and initiate star formation at high redshifts, which ends via percolation in the full reionization of the universe. This generates a UV and gamma background at high redshift ($z\sim20$) that could be seen today redshifted into the infrared and soft X-rays, respectively. The recent measurement of strong spatial correlations between fluctuations in the CIB and the diffuse X-ray background~\cite{Kashlinsky:2016sdv,Cappelluti:2017nww} suggests that a population of PBH could have initiated star formation and reionization at high redshift and also be responsible for the sources generating a fraction of the infrared and soft X-ray backgrounds today.  Intriguingly, the required abundance of PBH to explain these correlations is compatible with the opresent DM abundance~\cite{Kashlinsky:2016sdv}.  

{ \bf \textit{Hint 7}: The Chandra Deep Field South.} Recently, Chandra discovered more than 5000 SMBH in 1/6 sq.deg area in the Southern Sky~\cite{Luo:2016ojb}, in a deep image using more than seven million seconds exposure. This corresponds to over a billion SMBH in the entire sky, at distances up to 12 to 13 billion light years from us. Chandra found evidence that SMBH in the early universe (when it was just one billion years old) grow mostly in bursts, rather than by slow accumulation of matter. The seeds could be massive PBH that acquire mass very early (after photon recombination) via gas accretion and merging, and initiate a rapid growth of mass in bursts. These seeds will be responsible for an early and extended epoch of smooth (non-patchy) reionization, a scenario favored by Planck 2015 data~\cite{Heinrich:2018btc}.  Finally, these early SMBH could be precisely those seen by Chandra.

{\bf \textit{Other observational limits:}}  The preferred PBH mass distribution to explain the LIGO black hole masses and the detection of microlensing events in M31 and distant quasars passes easily all the other astrophysical constraints on the PBH abundance.   The non-disruption of wide binaries in the galactic halo~\cite{2014ApJ...790..159M} and the absence of radio and X-ray source correlations towards the galactic center~\cite{Gaggero:2016dpq} only constraints PBH mass distributions centered on masses above 10 $\Msun$ for $\sigma \simeq 0.5$.   Another stringent constraint comes from the matter accretion onto PBH in the early Universe that impacts the recombination history and affects the temperature fluctuations in the cosmic microwave background~\cite{Ali-Haimoud:2016mbv}.  Recent analysis~\cite{Poulin:2017bwe} suggests that PBH in the solar mass range composing the entirety of dark matter is disfavored.  However these constraints are very sensitive to the relative velocity between PBH and the baryonic gas (a power -4.8 dependence) and therefore CMB constraints can easily be accommodated if the PBH velocity reaches tens of km/s within PBH clusters.   It is nevertheless worth noticing that our scenario could be tested by future CMB experiments and observations of galactic center point sources.   

{\bf \textit{Conclusion and perspectives:} }   We have identified seven hints in favor of a PBH population with abundance comparable to dark matter, coming from various observations of widely different scales and epochs of the Universe or revealed by using physical arguments based on first principles.  These seven hints are namely: the rate and mass of BH binary mergers detected by AdvLIGO-Virgo; their spin distribution; the detection of microlensing of distant quasars and stars in M31;  the distribution and dynamics of faint dwarf galaxies and of their stellar clusters;  the evidences for cored DM profiles on different halo mass scales, from faint dwarfs to the brightest galaxies in clusters;  the observation of a population of SMBH very early in the history of the Universe; the spatial correlations of the source-substracted CIB and soft X-ray background fluctuations.   These clues suggest that the broad lognormal mass distribution of PBH is centered on $M_c \approx 3 M_\odot$ and has width $\sigma \approx 0.5$, a scenario allowed by the most recent constraints on PBH abundances and by the merger rates seen by AvdLIGO-Virgo.   Moreover, early matter accretion onto PBH in halos of mass $\lesssim 10^9 M_\odot$ and relaxation by two-body interactions in the densest environments naturally solve all the problems associated with the small-scale crisis of the standard cosmological model, the too-big-to-fail problem, the missing dwarf satellites, the core-cusp problem and the missing baryons.  PBH provide also the seeds to explain the existence of a population of SMBH at high redshifts.   It is worth noticing that each of these observations and problems could be solved individually by some other mechanism.  But it is intriguing that PBH Dark Matter can provide a common, natural and plausible origin to all of these, without requiring any specific parameter tuning.   

The present PBH Dark Matter scenario %of micro-clustered PBH 
opens a wide new field that is still vastly unexplored, and a series of observational perspectives that will either bring strong support or definitively rule out PBH as dark matter. 
%Among numerous perspectives, let us some of them are emphasized in the Appendix:  detecting a BH below the Chandraseckhar limit of $1.4 M_\odot$, the GW backgrounds and sources from PBH binaries or close encounters, the non-flat rotation curves of distant galaxies, hypervelocity stars in the Milky-Way halo, Low-Mass-X-ray-Binaries (LMXB), the 21cm signal from the cosmic dawn and reionization.  
%The PBH dark matter scenario opens a series of observational perspectives that will either bring strong support or definitively rule out PBH as dark matter. %Some of them are emphasized thereafter.
The most straightforward way to distinguish between stellar and primordial origins of BH binary mergers is to detect one involving a BH with a mass smaller than the Chandrasekhar limit of $1.4\,\Msun$.  This is within the target of upcoming runs of the AvdLIGO-Virgo Collaboration and our best scenario predicts that between 0.1\% and 1\% of the detected mergers will involve such a low-mass BH binary.  Improved statistics will also allow to better reconstruct the BH mass spectrum and spin distributions.

Another way to distinguish a primordial origin from preexisting stars is to detect the stochastic GW background from BH binaries, whose shape in frequency depends on the  clustering and formation mechanism of these binaries~\cite{Clesse:2016ajp}.   LISA~\cite{Bartolo:2016ami}, and eventually SKA pulsar timing arrays~\cite{Zhao:2013bba}, should have the required sensitivity to probe the different possible formation scenarios.  

Moreover, laser interferometers could also detect the bursts of GW coming from hyperbolic encounters of PBH in dense clusters~\cite{Garcia-Bellido:2017qal,Garcia-Bellido:2017knh}. These bursts have a well defined waveform and could be detected as coincident effects in all three detectors of LIGO-Virgo Collaboration, which would allow to distinguish them from spurious detector noise artefacts known as "tear-drop glitches". Those events would be complementary to the coalescing PBH binaries associated with bounded systems, and would allow us to determine the parameters of the clustered PBH scenario and break degeneracies present in the BHB events.

Furthermore, in a recent paper published in Nature, Genzel et al.~\cite{Genzel:2017jgd} found evidence that the rotation curves of distant galaxies, at redshifts $z=0.9 - 2.4$, were not flat, but actually decayed beyond twice the half-light radii $R_{1/2}$. This suggests that dark matter was subdominant in the outer regions of the disks of galaxies at those redshifts. Merger-tree hierarchical growth of galaxies suggests that the larger galactic halos grow by accretion of smaller halos. If DM is made of massive PBH, concentrated in dense clusters, as they merge with larger halos, they will tend to fall to the center, making halos "fatter" and increasing the rotation curves of galaxies. Some galaxies still exhibit decaying rotation curves at sufficiently large distances, like our own Milky Way~\cite{1604.01216} and Andromeda~\cite{Carignan:2006bm}, although to a much lesser degree that those early galaxies seen with KMOS.

Hypervelocity stars have been recently detected in the Milky Way halo and observations suggest a galactic bulge and LMC origin~\cite{2017MNRAS.469.2151B}  Those stars could have been slingshot out by a primordial IMBH~or by a massive black hole binary~\cite{Yu:2003hj}.  The existence of IMBH in the galactic center is supported by a recent observation~\cite{2017NatAs...1..709O}.   High-velocity stars are also detected in the core of globular clusters~\cite{Lutzgendorf:2012tn}, which might indicate an important population of massive PBH.  
 
The majority of Low-Mass-X-ray-Binaries (LMXB) have been detected in globular clusters or towards the galactic center, but not in the galactic disk, which favors a formation by tidal capture by the companion star.  In our PBH-DM scenario, one component of globular clusters and of the galactic bulge is the PBH population and we conjecture that it is dominant compared to BH of stellar origin and neutron stars.  Our scenario therefore suggests that some compact objects in LMXB having a mass below the Chandrasekhar mass are not neutron stars but PBH.  Because the amplitude of GW have a BH mass dependance and X-ray from LMXB do not, a broad PBH mass distribution centered on a few solar-mass would also explain why most LMXB have a mass in that range, whereas GW events involve heavier BH.

Other perspectives include the 21cm signal from the reionization that would be impacted by a PBH population~\cite{2013MNRAS.435.3001T,Gong:2017sie}, CMB spectral distortions~\cite{Nakama:2017xvq}, microlensing surveys towards the galactic center with Euclid~\cite{2011arXiv1110.3193L}, the monitoring of star position and velocities with GAIA. Heavy PBH from the queue of the mass distribution could also seed the formation of cosmic structures, as recently suggested in~\cite{Carr:2018rid}.  In particular we are exploring whether baryon accretion onto PBH and baryon-PBH gravitational scattering could induce an extra-cooling of the neutral hydrogen gas during the dark ages and cosmic dawn, and explain the unexpectedly large 21cm signal at redshift $z\approx 17$ detected recently by EDGES~\cite{2018Natur.555...38G}.

On the theory side, further investigations are certainly needed to refine the predictions of the PBH-DM scenario and to go beyond first principles, e.g. by developing N-black-holes simulations of the early Universe and of PBH halos (N-body simulations in which particles are individual BH with masses down to a few solar masses).   On the observational side, the existence of PBH in the mass range from 0.1 to 100 $M_\odot$ will be probed in the next few years by numerous independent observations, and any mass in this range should be covered by a least two independent probes.  Contrary to particle DM models, the PBH-DM hypothesis will therefore either be validated or soon ruled out by upcoming observations.   If clear evidence of PBH were found, it could initiate a paradigm shift on the nature of Dark Matter, with groundbreaking implications for our understanding of the early Universe and of High Energy Physics.

\vspace{1mm}
\section*{Acknowledgments} 
This work is supported by the Spanish Research Project FPA2015-68048-C3-3-P [MINECO-FEDER] and the Centro de Excelencia Severo Ochoa Program SEV-2012-0597. JGB thanks the Theory Department at CERN for their hospitality during a Sabbatical year at CERN. He also acknowledges support from the Salvador de Madariaga Program Ref. PRX17/00056.   The work of S.C. is supported by the Belgian Fund for Research FRS-FNRS through a \textit{Charg\'e de Recherche} grant.

\bibliography{biblio}

\appendix
%\section{Method} \label{sec:method}

\section{PBH merging rates} \label{sec:rates}

PBH merging  rates depend on the shape of the PBH mass spectrum.  Ref.~\cite{Bird:2016dcv} considered a monochromatic distribution of PBH masses but this is now clearly ruled out, so we have {\em extended} their model by considering a log-normal distribution of masses and kept the rest of their features.  We consider a PBH-DM model with a log-normal density distribution~\cite{Clesse:2015wea} of width $\sigma$, centered on the mass $\mu$,
\be
\psi \equiv \frac{ \dd \rho( \mPBH )}{\dd \log_{10} \mPBH} = \frac{\rho_{\rm DM}}{ \sqrt{2 \pi \sigma^2}}  \exp \left[ - \frac{ \log_{10}^2 \left( \mPBH/\mu  \right)}{2 \sigma^2}   \right],  \label{eq:PBHspectrum}
\ee
but the following results can easily be extended to any mass function and any PBH abundance.  Such a lognormal spectrum is motivated by PBH formation models in the context of inflation, e.g. in mild-waterfall hybrid inflation~\cite{Clesse:2015wea}.   The merger rates for the two considered clustering models are described thereafter:

\textit{1. Dominant clustering scale (DCS):}   The merger rate for PBH masses within $[m_{\rm A}, m_{\rm A} + \dd m_{\rm A}]$ and $[m_{\rm B}, m_{\rm B} + \dd m_{\rm B}]$, assuming the broad log-normal density spectrum of Eq.~(\ref{eq:PBHspectrum}), is given by~\cite{Clesse:2016vqa,Clesse:2016ajp}
\ba  \label{eq:rate1}
& \dfrac {\dd \tau_{\rm merg} }{\dd m_{\rm A} \dd m_{\rm B} }  \simeq  2.9 \times 10^{-9} ~ \langle \del^2 \rangle^{1/2}  \left\langle \left(\frac{v}{20~{\rm km/s}}\right)^{-11/7} \right\rangle \nonumber \\
& \times \dfrac{\psi(m_{\rm A}) \psi(m_{\rm B})  (m_{\rm A}+m_{\rm B} )^{10/7}  }{2^{10/7} \ln^2(10) \  \rho_{\rm DM} ^2 (m_{\rm A})^{12/7} (m_{\rm B})^{12/7}} \ \rr{yr}^{-1} \rr{Gpc}^{-3},
\ea
where $v$ is the relative PBH velocity.   We assumed $v = 20$~km/s, a typical value of the Virial velocity of faint dwarf galaxies. In the case of a monochromatic spectrum $m_A = m_B = 30 M_\odot$, the last factor is just 1, and one recovers the rate of Ref.~\cite{Clesse:2016vqa}.

\textit{2. Extended halo mass function (Ext-HMF):}   We have extended the calculation of~\cite{Bird:2016dcv} to compute the PBH merger to the case of a broad spectrum (\ref{eq:PBHspectrum}). One gets  
\ba    \label{eq:rate2}
&\dfrac {\dd \tau_{\rm merg} }{\dd m_{\rm A} \dd m_{\rm B} } \simeq  2~f_{\rm HMF} \frac{\rho_{\rm PBH}}{\rho_{\rm DM}} \left(\frac{M_{\rm c}}{400 M_\odot}\right)^{-11/21}   \nonumber \\
& \times \dfrac{\psi(m_{\rm A}) \psi(m_{\rm B})  (m_{\rm A}+m_{\rm B} )^{10/7}  }{ 2^{10/7} \ln^2(10) †\rho_{\rm DM} ^2 (m_{\rm A})^{12/7} (m_{\rm B})^{12/7}} \ \rr{yr}^{-1} \rr{Gpc}^{-3}.
\ea
   The parameter $f_{\rm HMF}$ depends on halo mass function ($f_{\rm HMF} \simeq 1$ for a Tinker HMF, $f_{\rm HMF} \simeq 0.6 $ for a Press-Schechter HMF~\cite{Bird:2016dcv}).   The critical halo mass  $ M_{\rm c}$ is the one below which halo evaporation takes place on a timescale less than about 3~Gyr.   In the case of PBH with $30~M_\odot$ masses, this critical mass was evaluated to be $M_{\rm c} \approx 400~\Msun$ in Ref.~\cite{Bird:2016dcv} but this value is subject to large uncertainties, e.g. related to possible halo concentrations.    To first approximation, the halo relaxation time scale is given by $t_{\rm relax} \approx R N / (\ln N \ v_{\rm vir}) $, where $N$, $R$ and $v_{\rm vir}$ are the number of PBH in the halo, the halo radius and its Virial velocity, respectively.   One therefore expects the merger rate to be boosted for lower PBH masses, e.g. by a factor three for PBH distributed all around $3~\Msun$.   On the other hand, for a broad distribution the total rate is also boosted, e.g. by two orders of magnitudes for $\sigma \approx 1$~\cite{Clesse:2016ajp}.  One should nevertheless consider that only PBH with masses larger than $\sim 1~\Msun$ could so far have contributed to the LIGO merger rate~\cite{Bird:2016dcv,Clesse:2016vqa}. 
  
Eqs.~(\ref{eq:rate1}) and~(\ref{eq:rate2}) have been integrated out numerically within the range $1 < m_{\rm PBH}/M_\odot < 100 $ to get the total expected rate for LIGO.  
These two models induce PBH merger rates consistent with the ones inferred by LIGO, from a few to a few hundreds events/year/Gpc$^3$.   

Contradictory results were obtained in Refs.~\cite{Sasaki:2016jop,Ali-Haimoud:2017rtz} by considering the lifetime of PBH binaries that are formed in the early Universe.  Assuming a monochromatic distribution, one can indeed get merger rates about two order of magnitudes larger than is allowed by LIGO.  However, for a broad distribution, it is more complicated to determine the orbital properties of PBH binaries, which do not only depend on the closest PBH with the same mass, but also on the heavier+farther and on the lighter+closer PBH from the tails of the distribution.  The first effect was considered in Ref.~\cite{Kocsis:2017yty} for not too wide distributions and was found not to reduce much the merger rate.  But on the other side, it is expected that a system of sub-solar mass PBH would form around heavier ones.   They would disturb the center of mass of the system, no longer located around the solar-mass PBH, which would contribute to reduce the eccentricity of massive PBH binaries formed in the early Universe.  These would then be more long-lived and thus have a suppressed merger rate.   In Ref.~\cite{Raidal:2017mfl} was also studied the stochastic gravitational wave background induced by a monochromatic and uniformly distributed in space distribution of PBH, following the scenario of Nakamura et al.~\cite{Nakamura:1997sm} for early binary formation. In this case, the PBH start at rest and only begin to approach each other when their relative distance is of order the horizon during radiation. Much later, during the matter era, three-body interaction with a more distant PBH will bind the two initially free PBH, inducing large eccentricities ($e\simeq1$) of the binary orbits, and thus generating a significant GW emission, which is proportional to $(1-e^2)^{-7/2}$. This produces a large GW background that is in conflict with LIGO measurements. In our scenario~\cite{Clesse:2016vqa,Garcia-Bellido:2017fdg,Clesse:2016ajp}, PBH are formed in micro-clusters, with a wide range of masses within the cluster, and decouple much earlier from the expansion. The distribution of velocities is very different from the previous scenario and the bound orbits rarely become very eccentric. Motion is mainly stochastic, with the influence of lighter/heavier PBH closer/farther from the binary are also relevant. A detailed analysis of this phenomenon was studied in Ref.~\cite{Sigurdsson:1993zrm} who did numerical N-body simulations of large numbers of black holes in globular clusters and found that very few binaries survived the stochastic tidal disruption of the other BH in the cluster. This means that the merger rate is suppressed and so is the stochastic gravitational wave background~\cite{Clesse:2016ajp}. More detailed investigations are certainly needed to determine unambiguously the merger rate of early binaries, in the broad-mass case, in order to compare with the LIGO bounds.

\section{MCMC reconstruction of PBH the mass function}    \label{sec:MCMC}
 Given the black hole progenitor masses $m_i$ of LIGO events and the inferred merger rate $\tau$, as well as their uncertainties ($\sigma_i$ and $\sigma_\tau$ respectively), the Bayesian posterior probability distribution of $\mu$ and $\sigma$ has been reconstructed by using a Markov-Chain Monte-Carlo (MCMC) method.   The likelihood function was decomposed in two parts, the one coming from the BH masses and the one from the merger rate:
\be
\mathcal L(\mu,\sigma) =  \mathcal L^{\rm BH } (\mu,\sigma | m_{i=1,\ldots 2 N_{\rm events}},\sigma_{i})  \times \mathcal L^{\rm rate} (\mu, \sigma |  \tau, \sigma_\tau)~.
\ee
The latter one $\mathcal L^{\rm rate}$ has been computed by using either Eq.~(\ref{eq:rate1}) or~(\ref{eq:rate2}) for each pair of $(\mu,\sigma)$ values, assuming a Gaussian prior distribution on $\log \tau$, centered on $\tau = 42 \ \rr{yr}^{-1} \rr{Gpc}^{-3} $ and of width $\sigma_{\log \tau} = 0.98$.   Our results are nevertheless rather insensitive to the exact values of these parameters.\footnote{Different confidence intervals were given by LIGO, depending on the considered mergers and on the rate reconstruction algorithm.} The first factor $\mathcal L^{\rm BH }$ is computed by using the Bayes' theorem and the LIGO event likelihood $\mathcal L^{\rm event}(m_{\rm A},m_{\rm B} | \mu, \sigma) $ taking into account that the range of LIGO increases with the BH masses like $\propto m_{\rr A} m_{\rr B} / (m_{\rm A} + m_{\rm B}) $.    
This likelihood has been implemented within the \texttt{MontePython} code~\cite{Audren:2012wb}, used to derive the posterior probability distribution of $\mu$ and $\sigma$ using the  Metropolis-Hastings algorithm.  

\section{Consistency between microlensing surveys}  \label{sec:microlensing}

EROS initially claimed that compact objects up to a few solar masses cannot account for the whole dark matter.  However, it has been recently shown that the EROS analysis is too conservative and that a more realistic treatment can reconcile it with the MACHO survey~\cite{Alcock:2000ph} that detected microlensing events suggesting between $5\%$ and $50\%$ of compact objects in the galactic halo in the range $[0.1-1]M_\odot$~\cite{Hawkins:2015uja}.   It was also pointed out  that different effects (dark matter profile, PBH velocity) can  affect these constraints.  Finally, as noticed in~\cite{Clesse:2015wea,Garcia-Bellido:2017xvr}, the LMC and SMC tiny projected area on the sky and their relatively close distance implies that the event rate might be strongly suppressed by the probability to find a PBH halo along the line of sight.   For instance, for a Press-Schechter mass function, about 80\% of the total dark matter resides in massive halos having a probability lower than $10^{-4}$ to be along the line of sight, whereas the other 20\%  
are either uniformly distributed in the galactic halo or clustered in abundant mini-halos.   For these reasons, we did not represent in Fig.~\ref{fig:spectrumreconstruction} the microlensing constraints from the Magellanic clouds. One should nevertheless notice that the MACHO constraints from~\cite{Alcock:2000ph} are consistent with the M31 events seen by AGAPE~\cite{CalchiNovati:2005cd} and those from quasars~\cite{Mediavilla:2017bok}.  Other surveys like MOA~\cite{2011Natur.473..349S} and OGLE~\cite{Wyrzykowski:2015ppa} have detected microlensing events in the direction of the galactic bulge, but they have not yet been analyzed in the context of the PBH scenario described here.  It is interesting to note however that the mass distribution of dark lenses found by OGLE~\cite{Wyrzykowski:2015ppa} in their Fig. 4 is very similar to the distribution of PBH described in Hint 1.

Finally, if PBH are regrouped into micro-clusters, they will evade present microlensing constraints because such micro-clusters would act as single massive objects~\cite{Garcia-Bellido:2017xvr}.  The microlensing events reported by MOA and OGLE would then correspond to the small fraction of PBH that have been slingshot away from the PBH micro-clusters and are spread uniformly in the halo. The detections of numerous additional microlensing events by present and future surveys like DES, LSST and Euclid, would be extremely valuable. For example, a specific microlensing survey towards the galactic center, where numerous PBH should reside, will help set more stringent limits on the different mass scales and thus on the possible PBH mass spectrum.  They will also be able to unambiguously confirm or exclude the existence of abundant compact objects in galactic halos.

\section{Accretion in dwarf galaxies}  \label{sec:accretion}

In order to estimate the remaining baryonic fraction in dwarf galaxies, and the resulting mass-to-light ratio, after an accretion episode by PBH, we have considered a simple \textit{spherical-cow} toy model, in which dwarf galaxies are modeled by homogeneous spherical halos of mass $M_{\rm halo} $, radius $R_{\rm halo}$ and volume $V_{\rm halo}$, with all the dark matter  made of massive PBH with a single mass $m_{\rm PBH}$.   The accreted mass by a single PBH is described by the Hoyle--Bondi accretion rate~\cite{Poulin:2017bwe},
\be
\dot m = 4 \pi \lambda \rho_{\infty}  v_{\rm eff}  r_{\rm HB}^2
\ee
where $\rho_{\infty} \equiv f_{\rm b} \rho_{\rm PBH}$ is the gas density ($f_{\rm b}$ denotes the baryon-to-PBH ratio) far from the PBH, and $\lambda$ is a parameter describing the accretion efficiency, whose maximal value is $\lambda \sim \mathcal O(1)$ for Eddington accretion.  The exact value of $\lambda$ is model dependent, and we followed here the arguments of~\cite{Poulin:2017bwe} suggesting that an accretion disk is generically formed\footnote{One can check straightforwardly that the arguments of~\cite{Poulin:2017bwe} for the early Universe remain valid in dwarf galaxies.} and we adopted the same value, $\lambda = 0.1$.  Moreover, $r_{\rm HB}$ is the Hoyle--Bondi radius,
\be
r_{\rm HB} \simeq 1.3 \times 10^{-4} {\rm pc } \, \left( \frac{m_{\rm PBH}}{M_\odot} \right)  \left( \frac{v_{\rm eff}}{5.7 \, {\rm km/s}} \right)^{-2}~,
\ee 
and $v_{\rm eff}$ is the PBH relative velocity with respect to the baryonic gas.  We calibrated $v_{\rm eff}$ to the Virial velocity inferred from observations of Eridanus II~\cite{Li:2016utv}, $v_{\rm eff}^{\rm EriII} = 6.9~{\rm km/s} $,  rescaled with the halo mass and radius according to the Virial theorem,
\be
v_{\rm eff} (M_{\rm halo}, R_{\rm halo}) = 6.9 \, {\rm km/s}  \left( \frac{M_{\rm halo} r_{1/2}^{\rm EriII}}{M_{\rm EriII} R_{\rm halo} }  \right)^{1/2}~,
\ee
in which the mass of Eridanus II within its half-light radius, $r_{1/2}^{\rm EriII} \approx 277 \, {\rm pc} $, is given by $M_{\rm EriII} \approx 1.2 \times 10^7 M_\odot$.   
 
Then, the total mass accreted by all the PBH in a single halo is given by $\dot m^{\rm tot} = \dot m  V_{\rm halo}   \rho_{\rm PBH}/ m_{\rm PBH} $ and the baryon-to-PBH ratio is governed by
\be
\dot f_{\rm b} = -  \frac{\dot m^{\rm tot} }{V_{\rm halo} \rho_{\rm PBH}} =  -  4 \pi \lambda f_{\rm b}  \frac{\rho_{\rm PBH}}{m_{\rm PBH}}  v_{\rm eff}  r_{\rm HB}^2~.
\ee
If one neglects the increase of the PBH density due to accretion, that is sub-dominant, one can integrate the previous equation and get
\be
f_{\rm b} (t) = f_{\rm b}^{\rm ini}  \exp \left[-  4 \pi \lambda f_{\rm b}  \frac{\rho_{\rm PBH}}{m_{\rm PBH}}  v_{\rm eff}  r_{\rm HB}^2 t \right]~.
\ee
Through the velocity dependence, this relation shows that PBH accretion provides a natural mechanism to explain that mass-to-light ratios are higher in denser, and thus fainter dwarf galaxies, thus providing a possible solution to the too-big-to-fail problem.   One can indeed notice that the accretion is the same for halos having the same density, whatever is their size or mass.   We have then assumed that the remaining fraction after one Gyr has entirely collapsed to form a dominant Population II stars such as the one observed in faint dwarfs.  The mass-to-light ratio is then  $f_{\rm b}^{-1} (1Gyr) $.  We have used the main sequence luminosity-mass relation, $L\propto m_{*}^{4.7} $ and assumed $m_{*} = 3 M_\odot $ to get a relation between the remaining baryon fraction and the absolute magnitude $M_{\rm V}$ of the galaxy,
\be
M_{\rm V} = 4.83 - 2.5 \log \left[ N_*  \left(†\frac{m_*}{M_\odot}\right)^{4.7}  \right]~,
\ee
where $N_* = f_{\rm b}(1 {\rm Gyr}) M_{\rm halo} / m_*$ is the number of stars in the dwarf.   Varying $m_*$ or the luminosity-mass relation simply shifts the absolute magnitude, and thus also the transition from luminous to faint objects. 

Despite a series of crude approximations, this calculation shows that it is \textit{in principle} possible that accretion onto PBH-DM plays a dominant role in the observed mass-to-light ratios of dwarf galaxies.   Important progresses in our understanding of matter accretion, PBH clustering, star formation,  are certainly needed to get a more accurate picture of this mechanism. The most important factor that determines the efficiency of this mechanism is certainly the PBH velocity distribution, because $\dot m \propto v_{\rm rel}^4$, itself depending on the PBH density profile and eventually on micro-clustering.  
N-body simulations accounting for gas accretion onto PBH and star formation could shed the light on the precise mechanism at play.  Other effects such as the DM profile, supernovae feedback, different stellar populations, etc. should also be considered and contribute to dwarf galaxy histories.  It is nevertheless remarkable that with such a simple toy model and realistic parameter values, one obtains mass-to-light ratios that are consistent with current observations.  

%\subsection{Perspectives}

\end{document}